\documentclass[useAMS,usenatbib]{mn2e}

\input epsf

\def\microlensing{\rm microlensing \;}
\def\inputs{\rm inputs}
\def\be{\begin{equation}}
\def\ee{\end{equation}}
\def\beq{\begin{eqnarray}}
\def\eeq{\end{eqnarray}}

\def\spose#1{\hbox to 0pt{#1\hss}}
\def\lta{\mathrel{\spose{\lower 3pt\hbox{$\sim$}} \raise
2.0pt\hbox{$<$}}}
\def\gta{\mathrel{\spose{\lower 3pt\hbox{$\sim$}} \raise
2.0pt\hbox{$>$}}}

\title[Lightcurve Classification]
      {Lightcurve Classification in Massive Variability Surveys I:
       Microlensing}

\author[V. Belokurov et al.]
       {Vasily Belokurov$^1$, N. Wyn Evans$^{1,2}$, Yann Le Du$^1$ \\
       $^1$ Theoretical Physics, Department of Physics, 1 Keble Road,
       Oxford, OX1 3NP, UK \\
       $^2$ Institute of Astronomy, Madingley Rd, Cambridge, CB3 0HA, UK}

\pagerange{\pageref{firstpage}--\pageref{lastpage}} \pubyear{2002}

\begin{document}

\label{firstpage}

\maketitle

%--------------------------------------------------------------------
\begin{abstract}
This paper exploits neural networks to provide a fast and automatic
way to classify lightcurves in massive photometric datasets. As an
example, we provide a working neural network that can distinguish
microlensing lightcurves from other forms of variability, such as
eruptive, pulsating, cataclysmic and eclipsing variable stars.  The
network has five input neurons, a hidden layer of five neurons and one
output neuron.  The five input variables for the network are extracted
by spectral analysis from the lightcurve datapoints and are optimised
for the identification of a single, symmetric, microlensing bump.  The
output of the network is the posterior probability of microlensing.

The committee of neural networks successfully passes tests on noisy
data taken by the MACHO collaboration.  When used to process $\sim
5000$ lightcurves on a typical tile towards the bulge, the network
cleanly identifies the single microlensing event.  When fed with a
sub-sample of 36 lightcurves identified by the MACHO collaboration as
microlensing, the network corroborates this verdict in the case of 27
events, but classifies the remaining 9 events as other forms of
variability. For some of these discrepant events, it looks as though
there are secondary bumps or the bump is noisy or not properly
contained. Neural networks naturally allow for the possibility of
novelty detection -- that is, new or unexpected phenomena which we may
want to follow up. The advantages of neural networks for microlensing
rate calculations, as well as the future developments of massive
variability surveys, are both briefly discussed.
\end{abstract}


\section{Introduction}

Variability in the sky has been known for thousands of years, but our
understanding of variable sources remains very incomplete. Some of the
most interesting objects in the sky are transient. These include
supernovae, microlensed stars, near-Earth or killer asteroids (which
are transient because of their exceptionally large proper motions)
optical flashes associated with gamma-ray bursts and stars undergoing
short-lived but key stages of stellar evolution like the helium core
flash and so on.  All these objects are rare. To hunt them down in a
systematic way means that we must record images, process the data in
real-time (or nearly so), recognise the events from their lightcurves
and archive them. 

The earliest examples of massive transient astronomy searches are the
microlensing surveys like
MACHO~\footnote{http://wwwmacho.anu.edu.au/},
EROS~\footnote{http://eros.in2p3.fr/} and
OGLE~\footnote{http://sirius.astrouw.edu.pl/\~{}ogle/}.
Typically, the surveys monitored $\sim 5 \times 10^6$ stars a few
times every night over several years in the directions of the Galactic
Bulge and the Magellanic Clouds, yielding $\sim 10^{10}$ photometric
measurements. Out of the $\sim 10^5$ sources which were variable, the
surveys tried to identify $\sim 10^2$ true microlensing events. The
selection criteria typically involved the imposition of sets of cuts
to ensure good lightcurve coverage and a steady baseline flux, to
require a single bump and thus eliminate common forms of stellar
variability and to require a good a statistical fit to the achromatic
standard microlensing lightcurve and so on. Many of the cuts developed
through trial and error, and evolved as the experiments progressed
(e.g., Alcock et al. 1997, 2000a). Unambiguous identification of
microlensing events was sometimes not possible, and the collaborations
sometimes reported their results in terms of two sets, one of high
quality events (any lightcurve that was undoubtedly microlensing) and
one of possible events (any lightcurve with a unique peak and a flat
baseline). Sometimes the cuts even eliminated interesting events --
for example, the longest ever microlensing event OGLE-1999-BUL-32 was
originally missed as its baseline flux was not constant and so failed
one of the imposed cuts (Mao et al. 2002).

Additionally, microlensing alert or early warning systems (e.g.,
Udalski et al. 1994) work by reducing the number of candidates to manageable
amounts. Each night's candidates are individually examined for the
onset of microlensing. Even for surveys as large as OGLE II, this
worked well. However, still larger surveys are planned for the future
and therefore it becomes important to automate the procedure and issue
alerts without human intervention.

The microlensing experiments are of course not the only massive
photometry searches being conducted by astronomers at the
moment. There are also collaborations primarily looking for supernovae
(e.g., The Supernovae Cosmology Project), optical flashes related to
gamma-ray bursts (ROTSE) and near-Earth asteroids (NEAT and
LINEAR). More generally, as Paczy\'nski (2001, 2002) has emphasised,
the monitoring of the optical sky for variability is likely to enjoy a
huge resurgence over the coming decade given the low cost of robotic
telescopes. The very near future will see terabyte datasets of
lightcurves routinely available to astronomers.  Such datasets will
contain complete samples of variable stars of all types, as well as
the very rare objects or events which primarily motivate the search.
It is a urgent and important problem to automate the classification of
lightcurves in massive variability surveys.

This paper argues that new analysis methods based on neural networks
will enable us to pinpoint and identify scarce transient objects in
such huge datasets. Our illustrative example is the identification of
scarce microlensing events against the background of variable stars.
However, we envisage that the applicability of the technique is much
wider.

% 2.
\section{Microlensing Lightcurves}

At any instant, the probability that a source star in the Galaxy shows
the microlensing effect is $\lta 10^{-6}$. Microlensing events are
hugely outnumbered by stellar variability which is at least a hundred
thousand times more common.  The lightcurve classification problem is
to devise algorithms that diagnose the different kinds of variability.
For applications to microlensing, the algorithm must distinguish
microlensing from other sources of variability (whether intrinsic
or extrinsic).

Let us assume a single, point-like, dark lens.  The microlensing
lightcurve has a characteristic form written down by Paczy\'nski
(1986).  The lightcurve is symmetric and achromatic. As the
probability of microlensing is so low, the variability must not
repeat. Microlensing is readily distinguished from some, but
unfortunately not all, forms of stellar variability. A cautionary
history is provide by the fate of the candidate event EROS-LMC-2.
This was one of the microlensing candidates uncovered by the
photographic plate search of the first phase of the EROS experiment
towards the Large Magellanic Cloud (Ansari et al. 1996). Although the
source star of EROS-LMC-2 was known to be variable at a low level
(Ansari et al. 1995), nonetheless microlensing seemed favoured by the
excellent fit of the lightcurve to the datapoints. However, there was
a substantial second bump in the lightcurve eight years after the
first, and EROS-LMC-2 was then discarded as a microlensing candidate
(Lasserre et al. 2000).

The background in microlensing databases is composed of periodic
variables (e.g., Cepheids, RR Lyrae), eruptive variables (e.g., dwarf
novae, classical novae), semi-regular variables (e.g., bumpers) and
the supernovae occurring in galaxies behind the source population.  Of
these, the most troublesome in microlensing surveys towards the
Magellanic Clouds and Andromeda are the bumpers and the novae-like
objects.  Although SNe Ia have reasonably well-understood lightcurves,
the same is not true of other types of supernovae which can mimic
microlensing rather well (for example, events 22 and 26 of Alcock. et
al. 2000a). Long period bumpers may be present as single bumps even in
5 seasons worth of data and they can be well-fit by the standard
microlensing lightcurve.

Let us stress that the identification of microlensing events remains
an awkward -- and not fully solved -- problem. For example, it
probably lies at the heart of the seeming discord between the results
of the MACHO and EROS experiments towards the Large Magellanic Cloud
(LMC). The MACHO group identified between 13 and 17 events towards the
LMC, whereas the competing EROS group found only 3 (Alcock et
al. 2000a, Lasserre et al. 2000). Although the exposure times and
field locations between the two experiments do vary, nonetheless the
rate found by MACHO is at least twice than that found by EROS.  This
same disparity is also seen in the experiments towards the Glactic
Bulge, as MACHO find an optical depth to microlensing of $\sim 3.23
\times 10^{-6}$ (Alcock et al.  2000b), whereas the EROS value is
about half of this (Afonso et al 2003).  Possible explanations are
that the MACHO selection algorithm may be too loose (causing
contamination with other variable sources), or that the EROS selection
algorithm may be too harsh (causing genuine events to be
discarded). It is here that neural networks may be able to make a
decisive contribution.

\begin{figure*}
\epsfxsize=12cm \centerline{\epsfbox{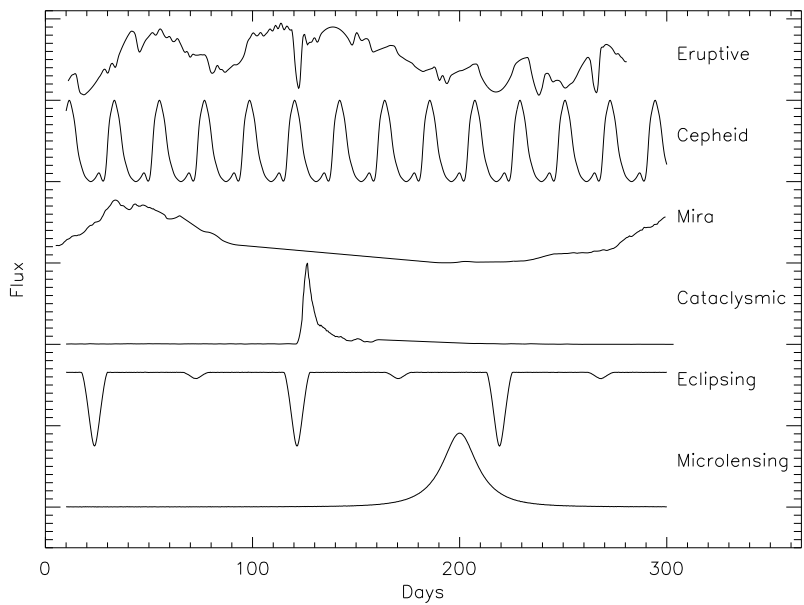}}
\caption{This shows sample lightcurves of different types of
fvariability included in the training and validation sets.}
\label{fig:examples}
\end{figure*}
\begin{table*}
\begin{center}
\begin{tabular}{l|c}\hline
Variable & Reference\\ \hline
Eruptive & van Genderen (1995), AAVSO \\
Puslating & Antonello \& Morelli (1996), AAVSO \\
Cataclysmic & Hamuy et al. (1996), AAVSO \\
Eclipsing & Brancewicz, Dworak (1980) \\
\hline
\end{tabular}
\end{center}
\caption{Sources of lightcurves of variable stars. AAVSO is the
American Association of Variable Star Observers.}
\label{table:eruptive}
\end{table*}

% 3
\section{An Informal Introduction to Neural Networks}

Neural networks have been used before for pattern recognition tasks in
physics (e.g., Bishop 1995). In particular, they are often used in
high energy physics experiments as triggers to select interesting
events from large datasets (M\"uller, Reinhardt \& Strickland 1995,
chapter 8).  Recent astronomical applications include classification
of optical stellar spectra (Bailer-Jones et al. 1997) and galaxy type
(Lahav et al. 1995), object detection in wide field imaging (Andreon
et al. 2000) and predictions of astronomical time series (e.g., Conway
1998, Perdang \& Serre 1998). There has also been a recent report of
preliminary results on automatic lightcurve classification by the
ROTSE collaboration (Wozniak et al. 2001). An interesting review of a
number of astronomical applications is in Storrie-Lombardi \& Lahav
(1994).

In a neural network, the neurons are arranged in layers. The input
data is fed to the bottommost layer. The output value emerges from the
topmost layer, the intervening layers are hidden.  The values of the
neurons in any layer $a_j$ are calculated via
\begin{equation}
a_j = \sum_{i} w_{ji} z_i.
\end{equation}
Here, $w_{ji}$ are the synaptic weights of the jth neuron with respect
to the ith neuron and $z_i$ are the activation values. The activation
value is computed from the value on the neuron via an activation
function $g$
\begin{equation}
z_i = g(a_i).
\end{equation}
As an activation function, we use the logistic function 
\begin{equation}
g(a) = {1 \over 1 + \exp(-a)},
\label{eq:logistic}
\end{equation}
which allows us to interpret the outputs of the network as {\it a
posteriori} probabilities (Bishop 1995, chapters 3,6).

We start with a sequence of input units (the ``patterns'') for which
the desired values of the output (the ``targets'') are known. This is
called the training set.  Given the patterns and a set of weights, we
can construct an error function $E$ which quantifies the performance
of the network. We want to obtain the weights $w_{ji}$ that minimise
the error function over the training set using a steepest descent
scheme.

We begin with random values for the weights and perform a sequence of
iterative up-dates using a variant of back-propagation as the learning
algorithm. The error derivatives with respect to the weights are
\begin{equation}
{\partial E^n \over \partial w_{ji}} = \delta^n_j z^n_i, \qquad\qquad
\delta_j \equiv {\partial E^n \over \partial a_j},
\end{equation}
where $n$ labels the pattern.  Using the chain rule, we obtain the
back-propagation formula
\begin{equation}
\delta^n_j = g^\prime( a_j) \sum_{k} w_{kj} \delta^n_k,
\end{equation}
which shows how the values of $\delta^n_j$ propagate through the
network, given the target value.  In each iteration, the weights are
up-dated according to the following rule
\begin{eqnarray}
\Delta w_{ij} = -\eta \sum_n\delta^n_j z^n_i,
\end{eqnarray}
where $\eta$ is the constant learning rate. The sum is performed over
all the patterns.  This is equivalent to the steepest descent method
of minimizing the error.  In practice, we use a refinement of this
algorithm, called resilient back-propagation, which helps to prevent
entrapment in local minima (see e.g., Bishop 1995, section 7.5.3).

As the network is converging to a minimum, it it important to prevent
overtraining. This is done by feeding a different set of patterns (the
``validation set'') to the network. The errors over the patterns in
the training and the validation sets are separately computed. The
training process is stopped just before the error in the validation
set begins to rise.  Finally, the performance of the fully trained
network can be assessed with a third set of patterns (the ``test
set''). It is important to ensure that the training, validation and
test sets do not contain any identical patterns.

% 4
\section{Implementation}

The experiments described below use the Stuttgart Neural Network
Simulator (``http://www-ra.informatik.uni-tuebingen.de/SNNS'').  Our
network is composed of one input layer, one hidden layer and one
output layer. The hidden layer is fully connected to the input and
output layers. There are 5 neurons in the input layer, 5 neurons in
the hidden layer and one neuron in the output layer.  The value of the
output neuron gives the probability that the event is
microlensing. The reason for the choice of 5 input neurons will become
obvious shortly.

\subsection{The Training and the Validation Sets}

There are three types of lightcurves in the training set -- simulated
microlensing events, variable star lightcurves from archival sources,
and sample lightcurves from a microlensing experiment (in this case,
the MACHO experiment).

Simulated microlensing events are generated by randomly choosing an
impact parameter, an Einstein crossing time between 7 days and 365
days and a time when the event reaches maximum. Random gaussian noise
is added with a dispersion in the range from 0.1 to 20 $\%$ of the
maximum flux. The lightcurves are sparsely sampled using the MACHO
sampling.

Variable stars may be divided into periodic variables and
eruptive/cataclysmic variables. The former are usually easier to
distinguish from microlensing than the latter, always provided more
than one period can be detected in the sampled datastream.  Examples
of typical lightcurves for different types of variability are shown in
Figure~\ref{fig:examples}. The periodic variables include pulsating
stars (such as Cepheids and Miras) and eclipsing stars.  Eruptive
variables include T Tauri, S Doradus and pre-main sequence
stars. Cataclysmic variables include novae, supernovae and symbiotic
variables.  The relative frequencies with which these stars occur are
not important in our analysis. All that matters is that the gamut of
shapes is well-represented in the training set. We are therefore
interested as much in regular representatives as in extreme examples
of the lightcurves.  Lightcurves for the variable stars are selected
from the sources listed in Table~\ref{table:eruptive}. For long data
sequences, the experimental window is placed randomly on the
lightcurve. In this way, we ensure that the bumps in the lightcurves
do not occur in a privileged place.

Finally, there are lightcurves randomly chosen from the MACHO database
(specifically, from field 113 towards the Bulge). The rationale for
this is that instrumental artefacts are certainly present in the MACHO
lightcurves and it is important for the neural network to be able to
recognise these.

The training set contains $400$ microlensing lightcurves, $150$
stellar variable lightcurves and $200$ MACHO lightcurves.  The
validation set contains the same number of lightcurves, although the
individual representatives are obviously different. The test set are
the $\sim 5 000$ lightcurves from MACHO tile 113.18292 which is part
of field 113 towards the Galactic bulge. Let us note that -- compared
with real data from a variability survey -- microlensing events are
over-represented in our training and validation sets.  The consequence
of this is that the network will provide more false positives (as the
prior probability of microlensing is too high). This is highly
desirable, as the best approach to detecting such an intrinsically
rare phenomenon as microlensing is to force fewer false negatives at
the expense of more false positives.

\subsection{Pre-Processing}

In many applications, it is both customary and advantageous to
pre-process data for feeding to the neural network. The main problem
with using raw photometry data is the curse of dimensionality (see
Bishop 1995, chapter 8). The simplest way of overcoming this is to
extract features of the lightcurve and use this as input to the
network. Properly implemented, this can lead to a very efficient network,
as prior knowledge can be incorporated and redundant variables can be
discarded in the pre-processing. However, there are dangers as well,
as important features in the lightcurves can be erased. 

The aim of a neural network is not to model the patterns but to model
the decision boundary between the patterns.  In microlensing surveys,
event identification normally proceeds by making sequences of cuts, in
which case the decision boundary is formed by a set of hyperplanes.
The advantage of a neural network over conventional sequences of
straight line cuts is that the former offers a better chance of
describing a complicated decision boundary accurately.

Microlensing events are characterised by the presence of a (iv)
single, (iii) symmetric, (ii) positive (i) excursion from the
baseline.  The event itself is characterised by (v) a timescale.
Motivated by these five features, we extract from the lightcurves the
following five parameters, which are inputs to the neural networks.

The first $x_1$ is the maximum value of the auto-correlation function.
This help to discriminate against noise and identify the presence of
any signal.  The second $x_2$ is calculated as follows. First, we
compute the median of the flux measurements which gives a good
approximation to the baseline. We then compute the mean of the
datapoints lying above and below the median and finally take their
ratio. This is then mapped on the interval [0.5,1] with the logistic
function. The input $x_2$ tests for the positiveness of the excursion.
The third $x_3$ is the maximum value of the cross-correlation function
of the lightcurve with the time-reversed lightcurve. This provides a
test for symmetric events. The fourth $x_4$ is the mean frequency
$\langle \nu \rangle $ calculated with the power spectrum $P(\nu)$ as
a weighting function.  For a periodic variable, we expect a shift in
the weighted mean frequency from zero. We compress $x_4$ with the
logistic function to lie in the range [$0.5, 1.0$].  Finally, the fifth
$x_5$ is the width of the autocorrelation function, as judged by its
standard deviation. If the event is microlensing, then the width is a
rough indication of the timescale.

\begin{figure*}
\begin{center}
\epsfxsize=12cm \centerline{\epsfbox{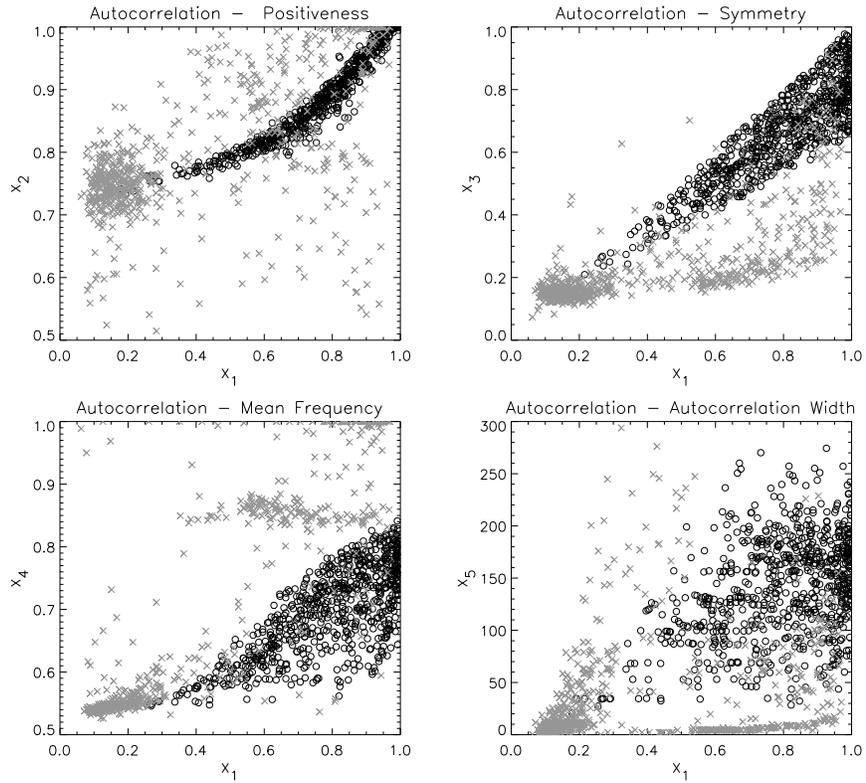}}
\end{center}
\caption{This shows projections onto the principal planes of the
five-dimensional space of inputs.  Bold circles show the microlensing
events and grey crosses the variable stars and noise in the training
and validation sets.}
\label{fig:inputs}
\end{figure*}

To motivate this choice of inputs, Figure~\ref{fig:inputs} shows the
locations of all the patterns in the validation and training sets. The
desideratum is that the choice of inputs offers a clear separation
between microlensing events and other patterns in the five dimensional
space ($x_1, \cdots, x_5$). The projections of this space onto the
principal planes offer grounds for believing this, as there is already
good partial separation in some of the plots (e.g., $x_1$ versus
$x_3$) and good evidence for regularities in others (e.g., $x_1$
versus $x_2$).  The final proof that the choice of inputs is good can,
however, only be provided by the performance of the network on the
test set.

Note that Figure~\ref{fig:inputs} plots unnormalized input variables;
however, the neural network uses normalized inputs. Scaling of the
inputs to numbers of the order of unity is often useful, as this
means that the network weights also typically take values of the same
order (Bishop, 1995, chapter 8). Pictorially, this can be thought of
as requiring the hyperplanes associated with each hidden unit to
intersect close to the origin and near the center of the datacloud.
For each input variable, this scaling is done by subtracting the mean
and dividing by the standard deviation to give the normalized inputs.

So far, we have skirted round the problem of missing datapoints. For
MACHO data, $\sim 10 \%$ of the lightcurves have gaps of the order of a
few days (aside from the 5 month gaps when the Galactic bulge is not
visible from Australia).  To compute the correlation functions, the
data is treated as if it were uniformly sampled. This gives rise to
some errors. If the typical gap size is much smaller than the event
timescale, then any errors we have introduced by this procedure will
be small. If the gap size relative to the timescale is very large,
then no classification can be plausibly extracted. If the gap size is
of the same order as the timescale, then the experiment needs
re-designing.  The input most sensitive to missing data is $x_4$
because this requires computation of the power spectrum. There are,
however, existing algorithms to do this for unevenly sampled data
(e.g, Lomb's periodogram as implemented by Press \& Rybicki (1989)
and Press et al. (1992)), which we employ.
 
Note that pre-processing gives rise to fast and powerful neural
networks, but it can also cause loss of potentially important
information in the data.  To check this, we can allow a neural network
itself to perform the projection. This leads to much bigger neural
networks which consequently take longer to converge. However, it does
have the advantage that no assumptions are built in from the
beginning. In this spirit, we experimented with a big neural network
which takes as the two input layers the unadulterated flux
measurements and errors at the sampling times and has $\sim 200$
hidden neurons. Once converged, the performance of this big network is
similar to the performance of smaller networks on pre-processed data.
From this, we draw the conclusion that our pre-processing has not
caused any serious degradation of information in the data.

\subsection{Training}

In training, the weights are initialised to random values. We then
perform iterations to reduce the error function
\begin{equation}
E^n = - \sum_{n} \left( t^n \log y^n + (1- t^n)\log (1- y^n) \right)
\label{eq:error}
\end{equation}
where $t^n$ and $y^n$ are the target and the response of the output
neuron for the $n$th pattern. We have chosen this form of the error
function (the so-called {\it cross-entropy} error function) as
appropriate for two class problems (see e.g., Bishop 1995, section
6.7). Given our choice of activation (\ref{eq:logistic}) and error
functions (\ref{eq:error}), the output $y^n$ approximates the
posterior probability $P(\microlensing |\inputs)$.

\begin{figure*}
\begin{center}
\epsfxsize=12cm \centerline{\epsfbox{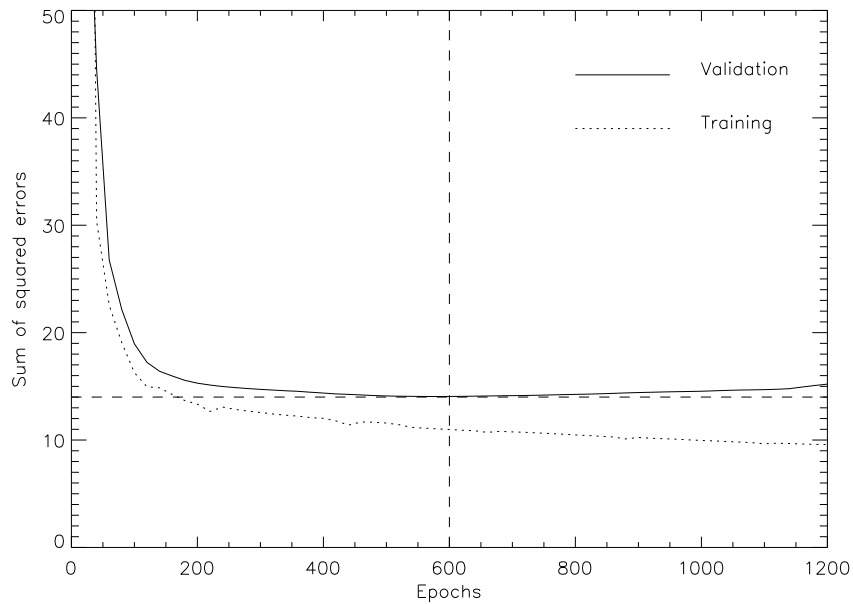}}
\end{center}
\caption{This shows the value of the cross-entropy error function
versus the epochs of training (number of iterations) for the patterns
in the training and validations sets. The long-dashed line shows the
point at which the training is stopped. The sum of errors is $\sim 15$
out of the 700 patterns in the set.}
\label{fig:earlystop}
\end{figure*}

The neural network must be able to generalise from the patterns in the
training set, and not merely reproduce them. A worry is that the
network will be over-trained and will reproduce structures of the
decision boundary in unnecessary detail. To guard against this, we use
early stopping as illustrated in Figure~\ref{fig:earlystop}. The
performance of the network on the validation set is compared to that
on the training set and the training stopped just before the error in
the validation set rises.  Another safeguard is provided by the
introduction of a small amount of noise to the weights on each
iteration, which guards against entrapment in a local minimum.

If training is started from different initial weights, we converge to
slightly different final weights. This makes it advantageous to use a
committee of 10 neural networks (Bishop 1995, section 9.6). There are
a total of 1500 lightcurves available. For each member of the
committee, the 1500 patterns are split in half randomly to give
validation and training sets with 750 members. For each pattern, the
final output is the average of the output of all 10 neural networks.

\begin{figure*}
\begin{center}
\epsfxsize=12cm \centerline{\epsfbox{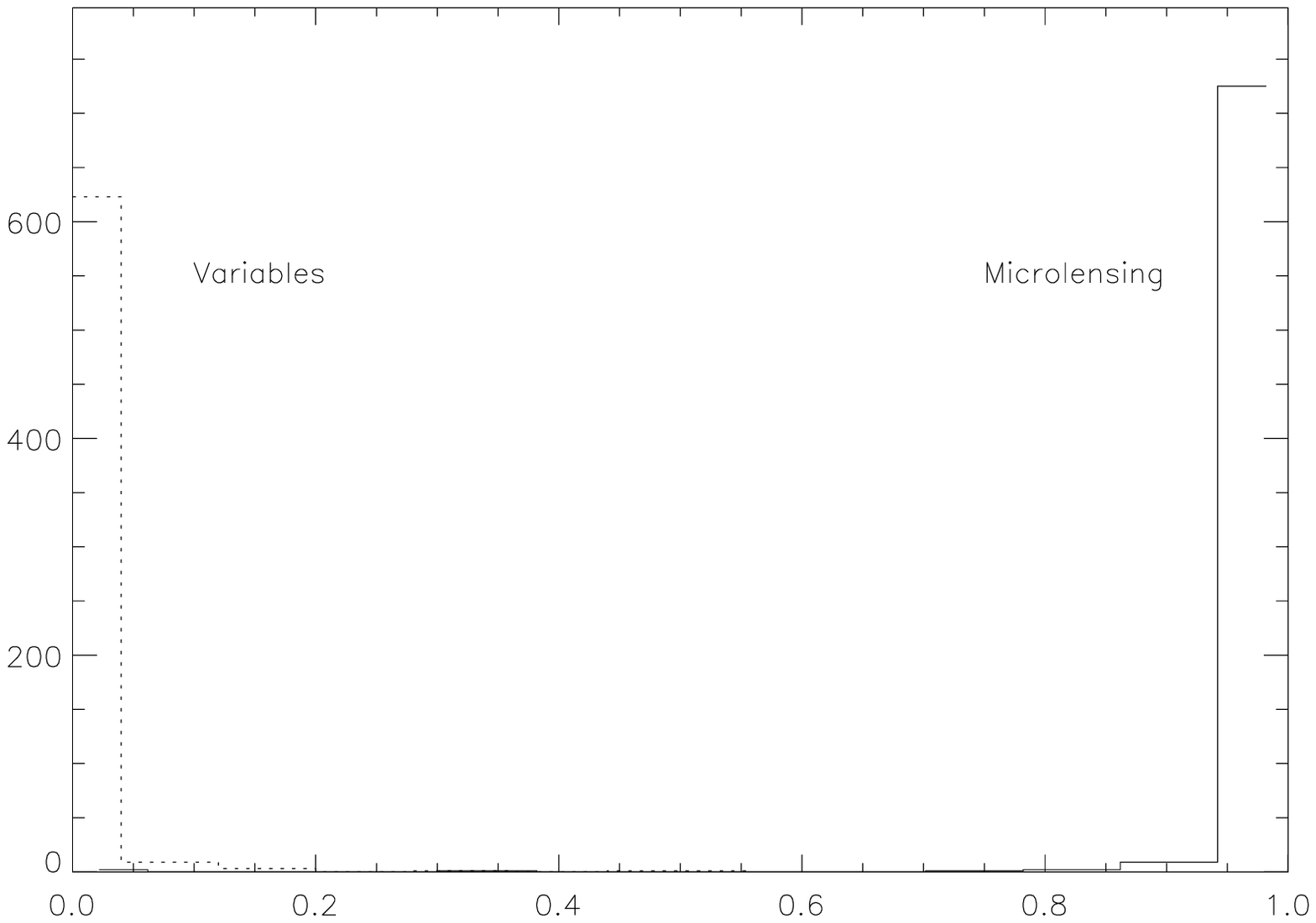}}
\end{center}
\caption{This shows the histogram of output values for the 1500
patterns in the validation and training sets. Note the clean separation
between microlensing and other types of variability.}
\label{fig:hist}
\end{figure*}

The histogram of the output values for the combined validation and
training set is shown in Figure~\ref{fig:hist}. There is a very clean
separation of microlensing events and other forms of variability.  The
non-microlensing events are strongly peaked at a probability $y=0$,
but there are a few events ($\sim 10$) that extend up to $y = 0.2$.
The microlensing events are strongly peaked at $y=1$, although again
there are a few ($\sim 10$) that extend down to $y = 0.7$.  The
probability $y = 0.5$ corresponds to the formal decision boundary
(Bishop 1995, section 10.3).  In fact, between $0.2 < y < 0.7$, there
are almost no events in the histogram. If, when presented with a
lightcurve, the neural network does give an output in this range, then
the classification is in reality uncertain. This is because any error
in the output can cause it to straddle the formal decision boundary.
This range of outputs really corresponds to patterns that are not
present in the training and validation sets. This is valuable as it
offers the possibility of the detection of unexpected and novel events
in variability surveys.

There are just 3 microlensing events out of 800 that are misclassified
(i.e., have $y < 0.5$). These are scarcely visible on the
histogram. It is interesting to locate these events in our input space
(see Figure~\ref{fig:inputs}).  These events have input coordinates
(-1.44, -1.07, -1.20, -1.17, -1.00), (-1.40, -1.16, -1.16, -1.14,
-0.8) and (-1.32, -1.11, -1.0, -1.11, -0.8). They are small amplitude
or short duration events dominated by noise, as indicated by the value
of the $x_1$ input which measures the presence of the signal.  There
is 1 false positive (i.e., a non-microlensing lightcurve with $y >
0.5$), which has coordinates ($-1.5, -0.7, -1.23, -1.2, -1.1$). This
is a lightcurve from the MACHO tile which is probably neither
microlensing nor variable star, but just noise.

\section{Tests towards the Bulge Fields}

\subsection{Normal Events}

\begin{figure*}
\begin{center}
\epsfxsize=12cm \centerline{\epsfbox{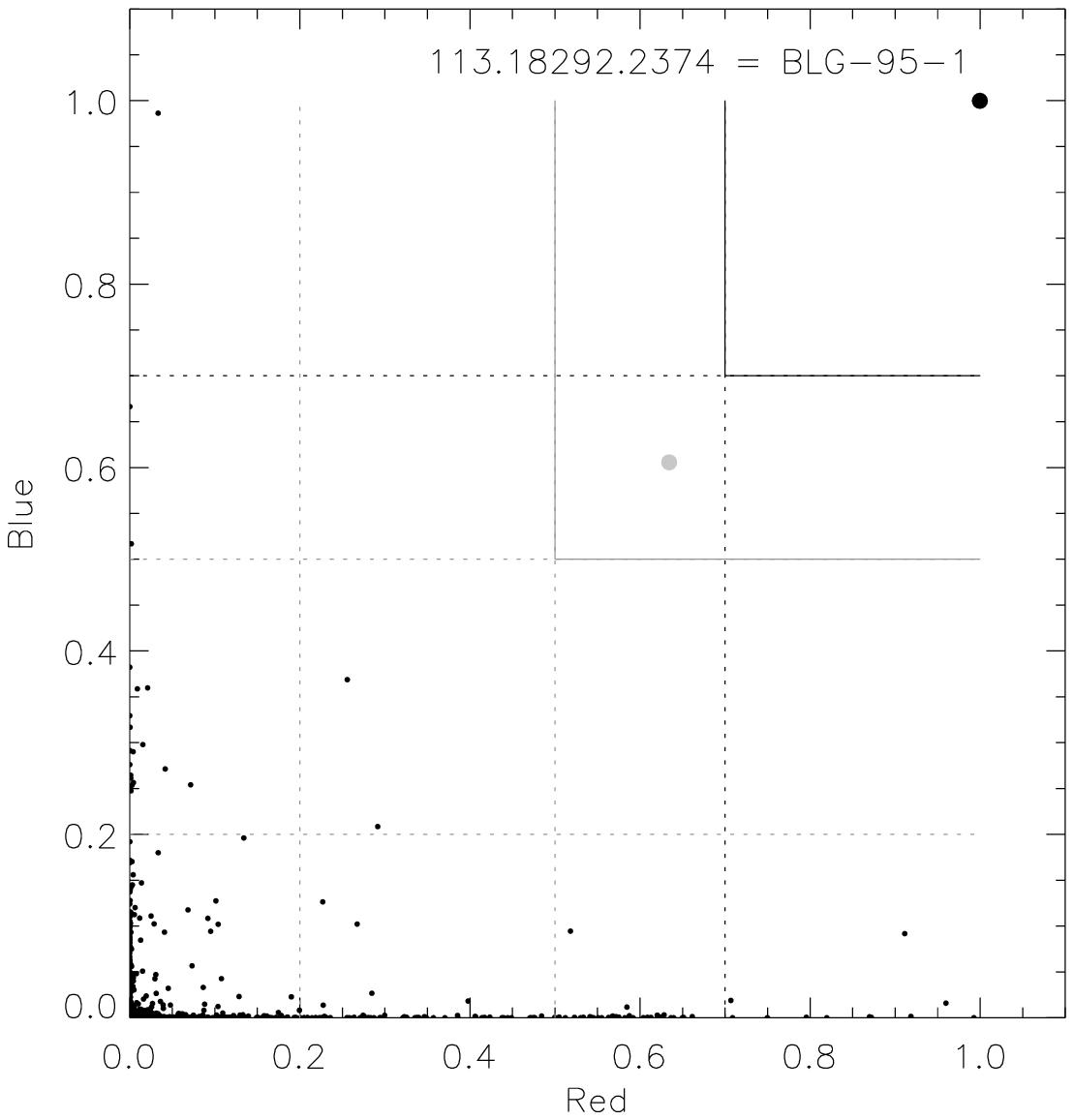}}
\end{center}
\caption{This shows the output of the committee of neural networks for
all lightcurves in tile 113.18292, which is publically available from
the MACHO project website (see Allsman \& Axelrod 2001). Shown on the
vertical and horizontal axes are the probabilities that the blue and
red lightcurves are microlensing. There are $\sim 5000$ lightcurves
on the tile, including one event BLG-95-1 identified by the MACHO
collaboration as microlensing. This is shown as the black spot.}
\label{fig:tile}
\end{figure*}
\begin{figure*}
\begin{center}
\epsfxsize=12cm \centerline{\epsfbox{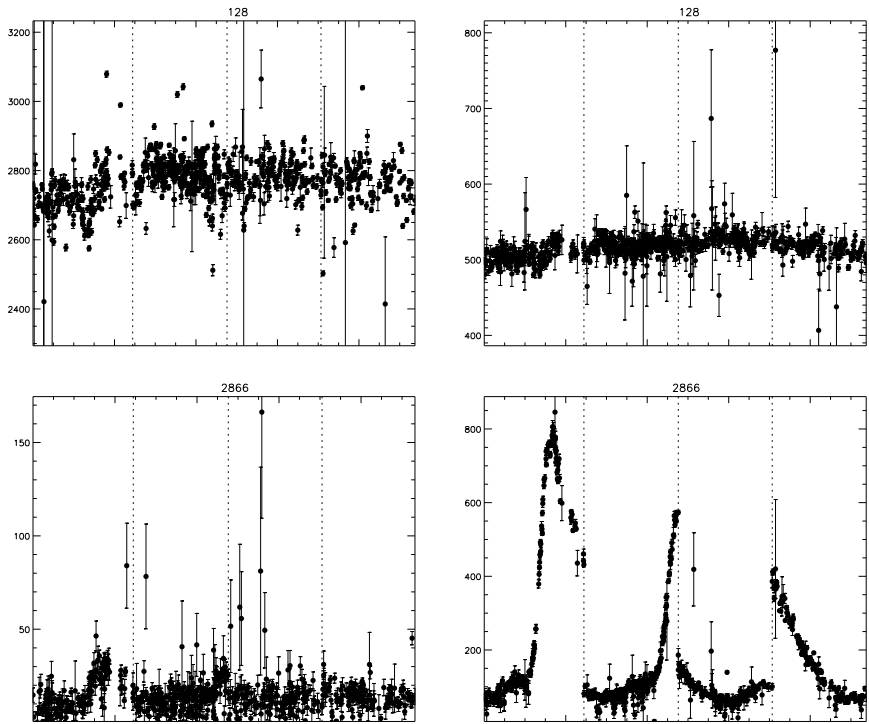}}
\end{center}
\caption{The upper panels show the blue and red lightcurves for the
event identified by the grey spot in Figure~\ref{fig:tile}. The lower
panels show the lightcurves for the event securely identified in blue
(left panel), but not in red (right panel). In all cases, the
horizontal axis is time in days, the vertical axis is flux in
ADU/s.}
\label{fig:pass}
\end{figure*}

\begin{figure*}
\begin{center}
\epsfxsize=16cm \centerline{\epsfbox{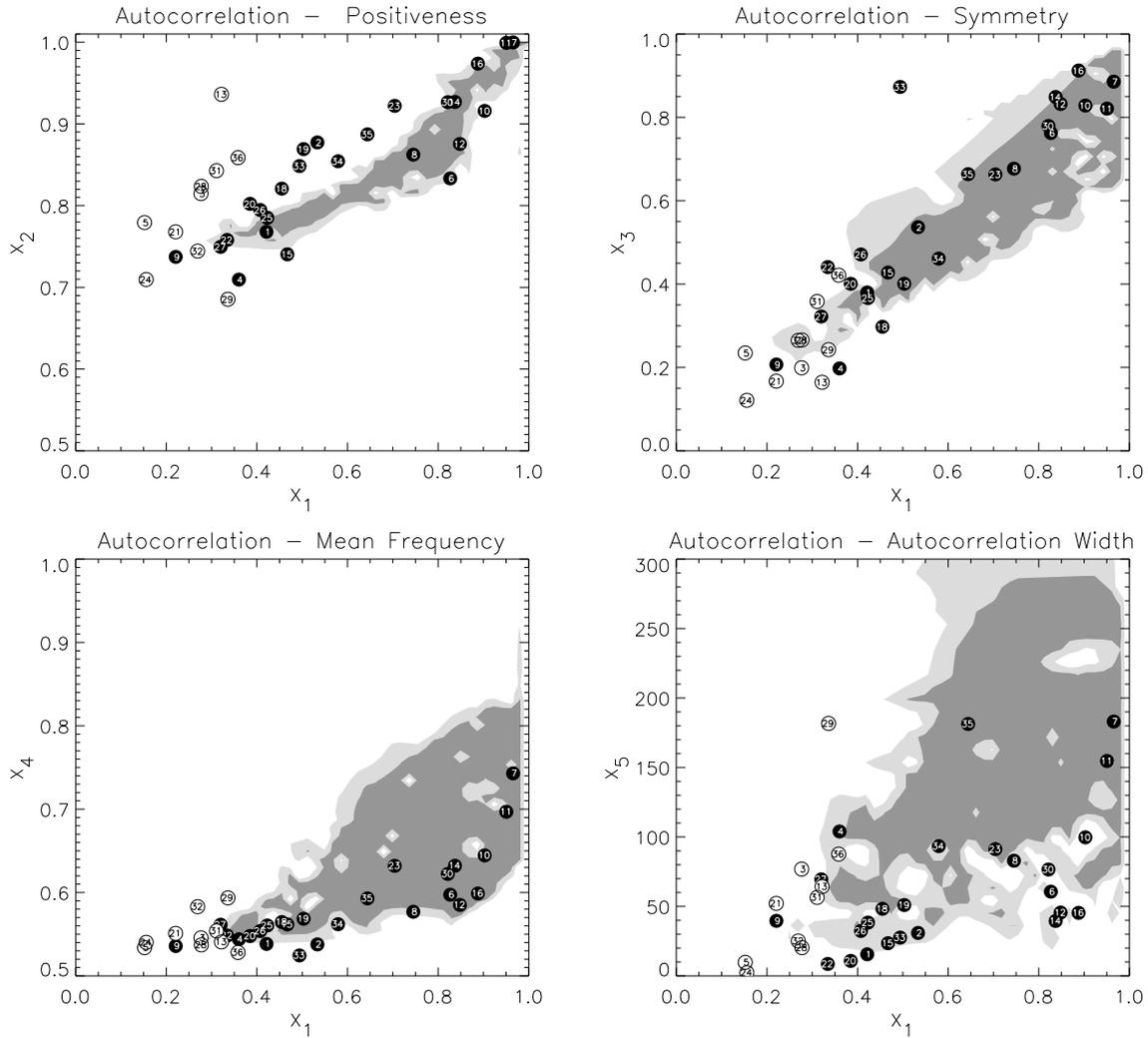}}
\end{center}
\caption{The grey-scale contours show the probability of microlensing
in the input space ($x_1, \cdots, x_5$) as judged from the patterns in
the training and validation sets. The circles show the locations of
the microlensing events identified by MACHO using conventional PSF
photometry. Filled circles designate the events also identified in the
red filter by the network. Unfilled circles are not
identified. Numbers within circles refer to our event designations in
Table~2. (Light grey means that the probability is greater than 0.5,
dark grey greater than 0.9).}
\label{fig:probconts}
\end{figure*}
\begin{figure*}
\begin{center}
\epsfxsize=12cm \centerline{\epsfbox{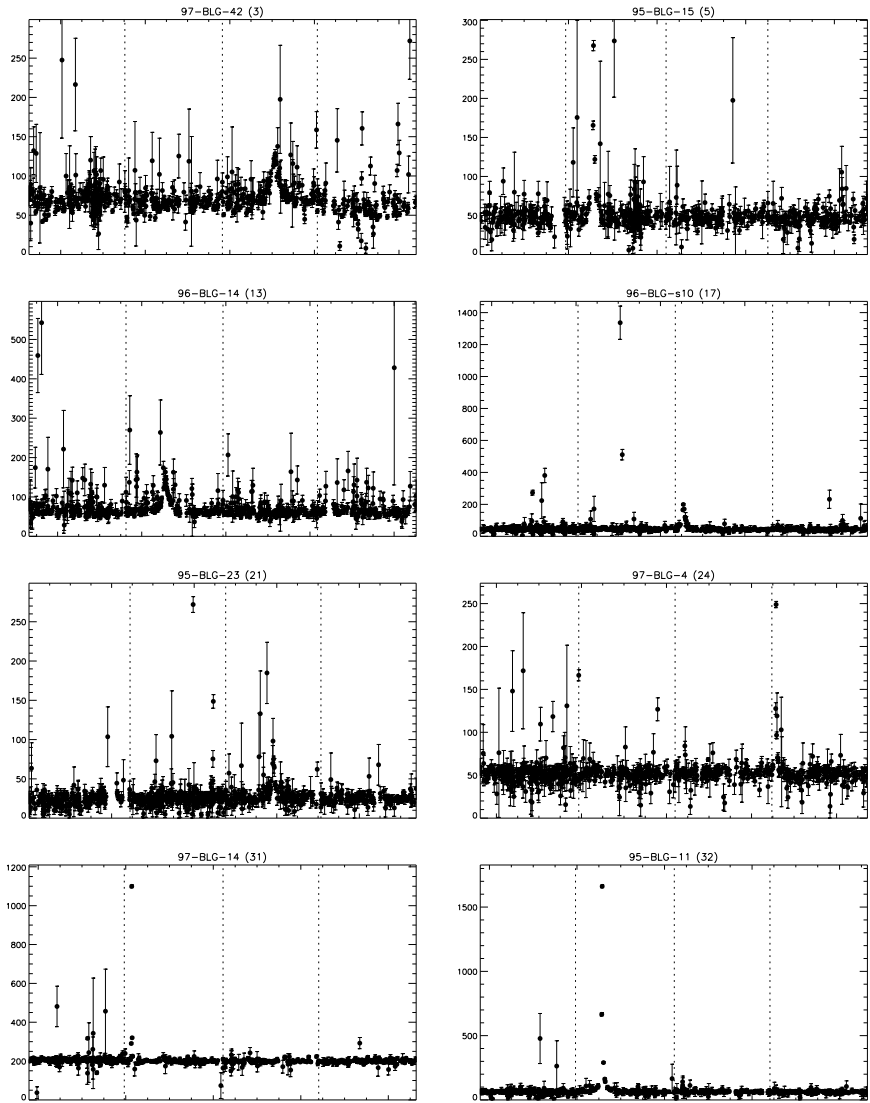}}
\end{center}
\caption{This shows the datapoints for eight of the events classified
as non-microlensing by the network and as microlensing by the MACHO
collaboration. The vertical axis is flux in ADU/s and the horizontal
axis is time in days. The data is presented as four strips of 7 month
sequences; the 5 months when the bulge is not visible from Australia
is marked by the vertical dashed lines.}
\label{fig:missing}
\end{figure*}
\begin{table*}
\begin{center}
\begin{tabular}{l|lccc||l|lcc}\hline
ID & MACHO ID & R &B & & ID & MACHO ID &R &B \\ \hline
 1 &  97-BLG-24 & 0.93 & 1.00 & \null & 
 2 &   95-BLG-5 & 0.92 & 0.96 \\
 3 &  97-BLG-42 & 0.00 & 0.00 & \null & 
 4 &  97-BLG-s4 & 0.72 & 0.00 \\
 5 &  95-BLG-15 & 0.00 & 0.00 & \null & 
 6 &  95-BLG-s8 & 1.00 & 1.00 \\
 7 &  97-BLG-18 & 1.00 & 1.00 & \null & 
 8 &  96-BLG-26 & 1.00 & 1.00 \\
 9 &  97-BLG-38 & 0.64 & 0.47 & \null & 
10 &  97-BLG-58 & 1.00 & 1.00 \\
11 &   96-BLG-1 & 1.00 & 1.00 & \null & 
12 &   97-BLG-2 & 1.00 & 1.00 \\
13 &  96-BLG-14 & 0.00 & 0.00 & \null & 
14 &  95-BLG-s9 & 0.99 & 0.91 \\
15 &  96-BLG-21 & 0.68 & 0.00 & \null & 
16 &   95-BLG-1 & 1.00 & 1.00 \\
17 & 96-BLG-s10 &    - & 0.16 & \null & 
18 &  96-BLG-20 & 0.99 & 1.00 \\
19 &  96-BLG-10 & 0.99 & 0.90 & \null & 
20 &   95-BLG-4 & 0.81 & 0.02 \\
21 &  95-BLG-23 & 0.00 & 0.00 & \null & 
22 & 95-BLG-s13 & 0.64 & 0.05 \\
23 &  95-BLG-10 & 1.00 & 1.00 & \null & 
24 &   97-BLG-4 & 0.00 & 0.00 \\
25 &  97-BLG-16 & 1.00 & 0.23 & \null & 
26 &   96-BLG-8 & 0.96 & 0.99 \\
27 & 95-OGLE-16 & 0.99 &    - & \null & 
28 &  95-BLG-39 & 0.16 & 1.00 \\
29 &   95-BLG-3 & 0.39 & 0.00 & \null & 
30 &  97-BLG-37 & 1.00 & 1.00 \\
31 &  97-BLG-14 & 0.21 & 0.00 & \null & 
32 &  95-BLG-11 & 0.01 & 0.00 \\
33 &  96-BLG-31 & 0.81 & 1.00 & \null & 
34 & 96-BLG-s16 & 1.00 & 0.83 \\
35 & 97-BLG-s14 & 0.80 & 0.90 & \null & 
36 &  95-BLG-22 & 0.30 & 0.79 \\ \hline
\end{tabular}
\end{center}
\caption{This shows the output of the committee of neural networks on
the subset of candidates towards the bulge in Alcock et al. (2000b)
which are selected on the basis of the conventional PSF photometry
package (SoDoPhot). The results of the analysis of the red and blue
lightcurves are shown separately. The output is the probability that
the event is microlensing. (Note that the red data for event 17 and
the blue data for event 27 are unavailable).}
\label{table:Macho}
\end{table*}

All MACHO lightcurves extracted with conventional PSF photometry (such
as SoDoPhot) are now publically available (Allsman \& Axelrod 2001).  As
a first test, we use lightcurves from tile 18292 of field number 113,
which lies towards the Galactic bulge. This tile contains $\sim 5000$
lightcurves of which one was identified by MACHO as a microlensing
event. The MACHO data are taken at a site with moderate
seeing. According to Alcock et al. (2000b), the median seeing is
$\approx 2.1$ arcsec.  This means that the quality of the data is
sometimes quite poor.  To allow for this, we clean the lightcurves by
removing all isolated points with more than $3 \sigma$ deviation from
the immediately preceding and succeeding datapoints. In general, this
makes good sense as it removes outliers, but it can sometimes remove
meaningful datapoints for very rapid brightness variations.

Each cleaned lightcurve is shown to the committee of neural networks.
The red and blue passband data are analysed separately.  In principle,
it would be advantageous to analyze the red and blue data together
because most variable stars show chromaticity differences. However,
this option is not open to us at the moment because the publically
available colour information on variable stars is still quite limited.
Figure~\ref{fig:tile} shows the results of the deliberations of the
committee. The probability of microlensing given the blue data is
shown against the probability given the red data. There is only one
pattern that satisfies this, namely the event identified by MACHO as
BLG-95-1. It is clearly and cleanly separated from the rest of the
patterns in the figure as a black circle in the topmost right
corner. There is an additional pattern that has output values $y
\approx 0.6$ for both the red and blue data.  This falls within the
regime of novelty detection.  Its input coordinates are
($-0.6,-1.5,-0.8,-0.66, 1.26$).  It is a very long event since $x_5 =
1.26$ is higher than typical values for microlensings. It falls into
poorly-sampled region in Figure~\ref{fig:inputs} which suggests why
this low signal-to-noise lightcurve was dragged into the microlensing
range.  Its lightcurve is shown in the upper panel of
Figure~\ref{fig:pass}. It is most probably a form of stellar
variability that does not lie in the training and validation sets. It
is interesting to note that there are a number of lightcurves with
output greater than 0.9 in one band, but not in the other.  Shown in
the lower panel of Figure~\ref{fig:pass} is a typical example, in this
case securely identified in blue ($y>0.95$) but not in red ($y <
0.05$). The blue lightcurve does indeed look like a microlensing
event, but the better sampling in the red passband shows a highly
active many-humped lightcurve which is most probably an eruptive
variable.

As a second test, we analyze the lightcurves for all 36 events in
Alcock et al (2000b) that were identified on the basis of conventional
PSF photometry. Table~\ref{table:Macho} shows the results of the poll
of the committee. In each case, the output of the neural network on
the red and the blue data is given. Of course, it is important to bear
in mind that the MACHO group's classification algorithm is itself
probably not 100 per cent efficient. There are reasons to believe --
both from the very high rate towards the Galactic Center that is
incompatible with theoretical models of the Galaxy and from the
differences between the MACHO and EROS results -- that the subsample
of candidates found by MACHO may have some contamination.  There are
total of 19 events identified with a probability $\gta 0.5$ as
microlensing in both the red and blue filters. In fact, these events
are all beyond reproach as microlensing candidates as the probability
$\gta 0.9$. Events 28 and 36 are securely identified in the blue data,
but the red data is corrupted.  Events 4, 9, 15, 20, 22 and 25 are
identified in the red data, but not in the blue.  Lastly, there are 9
events for which no microlensing signal whatsoever is detected (event
numbers 3, 5, 13, 17, 21, 24, 29, 31, 32). We shall examine the
lightcurves of some of these events shortly, but for the moment let us
emphasise that there is no guarantee that the original identification
by the MACHO collaboration was correct.

Fig~\ref{fig:probconts} shows the contours of probability for the
training and validation sets in the input space.  Light gray means
that the probability is greater than 0.5 and corresponds to the formal
decision boundary (see Bishop 1995, section 10.3). Dark gray means
that the probability is greater than 0.9 and corresponds to almost
certain microlensing.  The irregularity of the contours is due to the
fact that some regions are poorly sampled in the training and
validation sets. The contours have been drawn with a view to guiding
the eye.  Superposed on the contours in Fig~\ref{fig:probconts} are
the events.  The nine unfilled circles are those identified by the
network as variable stars but by MACHO as microlensing events. The
black circles are those for which both MACHO and the network agree as
microlensing.

There are a number of things to notice in the diagram. First, it is
evident that the network has the ability to extrapolate from the
validation and training sets and assign relative importance to the
combinations of features extracted by the input variables.  This is
clear because there are events securely identified although they lie
outside the contours (for example, event 18 is unambiguously
identified despite lying outside the probability contours in the
two top panels). Second, the $x_4$ input is the only one for which
explicit allowance has been made for noise and sampling. The network
seems to assign greater importance to this input, as almost all the
filled circles lie within the projected $90 \%$ probability
contour. This suggests that further improvements may be possible by
allowing for noise in the extraction of other input parameters (for
example, using extirpolation for the correlation analysis). Third, the
separation between the 0.5 and 0.9 probability contours is typically
very small, so the contour surface is very steeply rising. Such
outputs can correspond to novelty detection. Accordingly, they occupy
only a small region of the input space and so novelty detection occurs
-- as is highly desirable -- for only a few lightcurves. The small
separation between the contours provides justification for the sizes
of the training and validation sets. If there are too few patterns in
these sets, then the separation would widen. Such widening happens in
our network only in a few unimportant regions, which are physically
inaccessible (that is, such a combination of input variables gives
rise to lightcurves that do not occur in nature).  Fourthly , all the
unfilled circles have $x_1 < -0.8$ and so lie in the noise-dominated
regime. However, the values of $x_2$ indicate the presence of
substantial positive excursions. This is already enough to tell us
that the noise in the MACHO data is strongly non-Gaussian.

Figure~\ref{fig:missing} shows the lightcurves for 8 of the events
corresponding to the unfilled circles. For some of these events, it
looks as though there are secondary bumps (e.g., event 3). For others,
the bump is not properly contained (e.g., events 5 and 31) or the bump
is overwhelmed by noisy data (e.g., events 13 and 17). It seems that
the performance of our network is excellent, as these events certainly
need to be looked at with care before accepting a classification as
microlensing.  However, it is premature to conclude that MACHO have
misclassified these events. This is because the MACHO group have
re-processed all the lightcurves with difference image analysis (DIA)
and this will improve the quality of the lightcurves, reducing noise
and contamination from nearby stars.  However, without having the DIA
lightcurves, we cannot confirm their verdict of microlensing.

\begin{table}
\begin{center}
\begin{tabular}{l|lccc}\hline
ID & MACHO ID & deviation & R &B \\ \hline
 1 &  95-BLG-30 & f & 1.00 & 1.00 \\
 2 &  96-BLG-12 & p & 1.00 & 0.99 \\
 3 &  97-BLG-1  & b & 0.95 & 1.00 \\
 4 &  97-BLG-8  & p & 1.00 & 1.00 \\
 5 &  97-BLG-26  & p & 1.00 & 1.00 \\
 6 &  96-BLG-3  & b & 0.83 & 0.51 \\
 7 &  95-BLG-18  & p & 0.99 & 0.75 \\ \hline\end{tabular}
\end{center}
\caption{This shows the output of the committee of neural networks on
the exotic events identified towards the bulge in Alcock et
al. (2000b).  These are all exotic events selected on the basis of the
conventional PSF photometry package (SoDoPhot); f stands for
deviations due to finite source size, p due to parallactic effects and
b due to binarity. The output is the probability that the red and blue
data correspond to a microlensing event.}
\label{table:exotic}
\end{table}

\subsection{Exotic Events}

Some microlensing lightcurves can show deviations from the standard
Paczy\'nski form caused by parallactic or finite-source size effects
or by binarity and so on (see e.g., Mao \& Paczy\'nski 1993, Mao \& Di
Stefano 1995, Kerins \& Evans 1999, Mao et al. 2002).  In
Table~\ref{table:exotic}, all the exotic events identified in Alcock
et al. (2000b) using the SoDoPhot photometry package are processed
with the committee of neural networks.

Parallactic events (like 96-BLG-12) occur when the Einstein radius
projected onto the observer's plane is of the order of an astronomical
unit. In such a circumstance, the changing motion of the Earth around
the Sun during the event is detectable by an asymmetry in the
lightcurve with respect to the peak.  Events showing deviations caused
by finite source size (like 95-BLG-30) occur whenever the angular size
of the source is of the same order of magnitude as the angular
Einstein radius.  They are usually flatter-topped than the classical
Paczy\'nski curves for microlensing by a point source.  For both these
kinds of deviation, the committee of neural network performs well, as
shown in Table~\ref{table:exotic}. All the parallactic and finite
source size events are identified as microlensing.

However, binarity can cause much substantial deviations. For example,
strong binary events have additional peaks, although these can
sometimes be missed if sampled irregularly. Weak binary events may
just have distortions to the peak or the wings of the lightcurve.
Accordingly, we might expect the detection of binary lightcurves to
require the training and testing of a new neural network. This is
supported by the results in Table~\ref{table:exotic}.  Here, 96-BLG-12
is identified by the committee, whereas 96-BLG-3 falls into the domain
of novelty detection. It is reassuring that in the former case, the
event is recognised, while in the latter case, the event is recognised
as a new phenomenon. The development of software to recognise binary
events is a problem that has not been fully solved by any of the
microlensing collaborations to date. It seems reasonable to expect
neural networks to play a powerful role here.

\section{Conclusions}

This paper has devised a working neural network that can distinguish
simple microlensing lightcurves from other forms of variability, such
as eruptive, pulsating, cataclysmic and eclipsing variables.  The
network is structured to have five input neurons and one output
neuron. The inputs and output are separated by a layer of hidden
neurons. The simplicity of the network means that it can be trained
very quickly and it can be used to process huge datasets in less than
a second. Each lightcurve is pre-processed to provide five inputs to
be fed to the network. In our application, the five inputs were chosen
on physical grounds as good discriminants for microlensing.  In other
applications, different input variables may be optimum. Our network
has been constructed so that the output is the posterior probability
of microlensing.

We believe that neural networks offer three important advantages over
conventional techniques using in microlensing experiments.  First, the
decision boundary separating microlensing from non-microlensing may be
rather complicated. At present, all microlensing collaborations use a
series of cuts (for example, on the goodness of fit to a Paczy\'nski
curve, on achromaticity and so on).  This is the crudest form of the
decision boundary. However, even simple neural networks can reproduce
complicated decision boundaries and so the technique is both more
efficient and more flexible.  Moreover, once a lightcurve has failed
to pass a cut at the early stages of a conventional selection process,
it is lost for any further analysis. But, neural networks assign
relative importance to the input parameters, thus the decision is
based on the whole of the information available.

Second, neural networks offer a superior way of calculating the event
rate avoiding the need for any kind of efficiency calculation. The
classical procedure of identifying events with cuts is inefficient,
and this necessitates the cumbersome Monte Carlo calculation of the
numbers of synthetic events passing the cuts. However, a
properly-designed neural network can reproduce the decision boundary
well and can enable the event rate to be computed directly for
comparison with theoretical models, thus completely sidestepping the
need for any Monte Carlo calculation of the efficiencies.

Third, novelty detection is made both more precise and easier by
neural networks. The conventional approach relies on examination by
eye of the events left over after applying a sequence of cuts.  For
our neural network, we have argued that all lightcurves with outputs
between $0.2$ and $0.7$ may be examples of lightcurves not contained
within the training set. These are the events which need looking at
very carefully.  In the even more massive datasets of the future, it
will be important to identify possible novel events as quickly and as
efficiently as possible.

From the point of view of microlensing, it is interesting to extend
the work in this paper to include additional effects. Some of the
ongoing microlensing experiments are working in the highly blended
r\'egime. For example, the POINT-AGAPE (Paulin-Henriksson et al.
2002a,b), WeCAPP (Riffeser et al. 2001) and MEGA (Crotts et al. 2000)
collaborations are all monitoring the nearby galaxy M31. Here, the
individual stars are not resolved, so the flux in a pixel or
superpixel is followed (Baillon et al. 1993). The range of lightcurves
in such pixel lensing experiments is very wide -- for example,
microlensing events can occur in the same superpixel as bright
variable stars (e.g., the event PA-99-N1 described in Auri\`ere et al
2001). So, the identification of microlensing events becomes still
more daunting. As the complexity of the pattern recognition task
increases, so we expect the power and flexibility of the neural
network approach to pay increasing dividends.  Also, in this paper, we
have concentrated on the microlensing datasets towards the bulge, for
which the source stars are often bright. It is important to apply our
techniques to the microlensing events towards the Large Magellanic
Cloud. Here, the task is harder as the source stars are fainter and
there is serious contamination from supernovae in background
galaxies. This work will be the subject of a separate publication.

Although our application has been strongly focused on microlensing,
the technique is of general applicability in astronomy. There are
numerous ongoing or planned massive photometry surveys using robotic
telescopes (ROTSE), wide field cameras (WASP and VISTA) and
space-borne satellites (GAIA and Eddington). Although the goal of the
surveys is different, the basic method is the same -- brute force
search through many terabytes of data for interesting but rare events,
whether planetary transits, cataclysmic variables or optical flashes.
We envisage such tasks being routinely devolved to neural networks in
the astronomy of the future.  In each case, cascades of neural
networks could be trained to filter and identify the various classes
of variable stars, to pinpoint the target events of interest and to
isolate the unexpected or new classes of phenomenon which need looking
at very carefully.

\section{Speculations}

Suppose the goal is to monitor the whole sky for variability at short
time intervals down to 20th magnitude (roughly a billion objects in
our Galaxy). In this speculative final section, we ask what is
possible now and what will be possible by 2010?

Let us consider the simple situation of a single neural network
program running on a single computer.  The middle-range hardware
situation today is typically a processor running at 2200 Mhz
(corresponding to approximately 1000 MIPS or Million Instructions per
Second). In order to predict the situation in 2010, we can use
``Moore's Law", which says that the numbers of transistors in a
processor chip doubles every year or so. Thus, by 2010 the processor
speed should be $\sim 100\,000$ MIPS, compared to about 1000 MIPS
today. But, to evaluate the progress in run time, we must consider
both hardware and the compiler. Benchmarking programs such as SPEC
provide us with some clues as to what will be achieved in 2010. If we
look at the evolution in performance results on SPEC tests for
computers between ~1995 and ~2000, we find an approximate speed-up
factor of 16, or roughly 1.74 per year. We can extrapolate this
progress over the 2002-2010 period, which gives a speed-up factor of
~85. In other words, both Moore's Law and the extrapolation of
benchmarking suggest rather similar speed-up factors of roughly two
orders of magnitude by 2010.

The time required to run the neural network itself is negligible
compared to the time required to run the pre-processing, which
extracts the parameters used by the neural network. Our present
pre-processing program requires $10^{-4}$ s to analyse 100 data points
for a single star. We have chosen 100 datapoints as it might
correspond to sampling 3 times a night for one month, which is
reasonable for the detection of fast transient events.  Alternatively,
it might correspond to sampling once a night for 3 months, which is
reasonable for the detection of variability like microlensing with a
characteristic timescale of $\sim 1$ month.  At present, it therefore
takes $\sim 10^5$ s (or over a day) to analyse such dataset for the
whole sky.  By 2010, it will take only 20 minutes for such a program
to run on the whole sky (using the speed-up factor of 85).  More
generally, the time taken in seconds to analyse a set of $N_{\rm pts}$
data points for $N_\star$ stars in 2010 is
\begin{equation}
t \sim 2 \times 10^{-9} N_\star N_{\rm pts} \log N_{\rm pts}.
\end{equation}
Let us assume there are 8 hours of observing time a night and that we
wish to process a months data for the whole sky in real-time. Then we
can derive the real-time equation
\begin{equation}
t^2 \sim 1.7 \times 10^{6} \left( 13.7 - \log t \right).
\end{equation}
which has a solution $t \sim 50$ minutes. In other words, real-time
processing of variable phenomena across the entire sky down to 20th
magnitude will be possible for sampling rates of $\gta 1$ hr by 2010.

Our speculative calculation errs on the pessimistic side because we
have not taken into account any correction for application of parallel
processing or the fast developing GRID technology for high performance
computing.  However, it surely does enough to convince the reader
that, properly trained, neural networks can analyse huge datasets very
quickly. This will become one of the methods of choice for data-mining
in the massive variability surveys of the very near future.

\section*{Acknowledgments}
VB is supported by a Dulverton Scholarship.  YLD thanks PPARC, while
NWE thanks the Royal Society for financial support. In this research,
we have used, and acknowledge with thanks, data from AAVSO
International Database based on observations submitted to the AAVSO by
variable star observers worldwide.  This paper also utilizes public
domain data obtained by the MACHO Project, jointly funded by the US
Department of Energy through the University of California, Lawrence
Livermore National Laboratory under contract No. W-7405-Eng-48, by the
National Science Foundation through the Center for Particle
Astrophysics of the University of California under cooperative
agreement AST-8809616, and by the Mount Stromlo and Siding Spring
Observatory, part of the Australian National University.  In this
respect, we particularly wish to thank Robyn Allsman and Tim Axelrod
for help in acquiring data for an entire tile. We also thank the
anonymous referee for a helpful report.

\label{lastpage}

\end{document}